\documentclass[aps,prl,reprint,twocolumn,superscriptaddress,floatfix,nofootinbib,longbibliography]{revtex4-1}
\usepackage{graphicx,amsmath,amsfonts,amssymb,amsthm,xr}
\usepackage{epsfig,amsmath,amssymb,color,dsfont,upgreek,physics}
\usepackage{mathrsfs}
\usepackage{widetable}
\usepackage{mathtools}
\usepackage{bbold}
\usepackage{float}
\usepackage{graphicx}
\usepackage{dcolumn}
\usepackage{bm}
\usepackage{xcolor}
\usepackage{braket}
\usepackage{hyperref}
\usepackage[final]{changes}

\AtBeginDocument{\let\oldcontentsline\contentsline}
\newcommand{\notoccontentsline}[4]{\oldcontentsline{}{}{}{}}
\newcommand{\droptocpage}{\addtocontents{toc}{\let\protect\contentsline\protect\notoccontentsline}}
\newcommand{\incltocpage}{\addtocontents{toc}{\let\protect\contentsline\protect\oldcontentsline}}

\definecolor{myblues}{cmyk}{073,0.00,0.09,0.38}

\definechangesauthor[name={ZS},color=myblues]{ZS}

\begin{document}

\preprint{APS/123-QED}

\title{An Orbit-qubit Quantum Processor of Ultracold Atoms}

\author{Ming-Gen He}%
\thanks{M.-G.H,~and W.-Y.Z. contributed equally to this work.}
\affiliation{Hefei National Research Center for Physical Sciences at the Microscale and School of Physical Sciences, University of Science and Technology of China, Hefei 230026, China}
\affiliation{CAS Center for Excellence in Quantum Information and Quantum Physics, University of Science and Technology of China, Hefei 230026, China}

\author{Wei-Yong Zhang}%
\thanks{M.-G.H,~and W.-Y.Z. contributed equally to this work.}
\affiliation{Hefei National Research Center for Physical Sciences at the Microscale and School of Physical Sciences, University of Science and Technology of China, Hefei 230026, China}
\affiliation{CAS Center for Excellence in Quantum Information and Quantum Physics, University of Science and Technology of China, Hefei 230026, China}

\author{Zhen-Sheng Yuan}%
\affiliation{Hefei National Research Center for Physical Sciences at the Microscale and School of Physical Sciences, University of Science and Technology of China, Hefei 230026, China}
\affiliation{CAS Center for Excellence in Quantum Information and Quantum Physics, University of Science and Technology of China, Hefei 230026, China}
\affiliation{Hefei National Laboratory, University of Science and Technology of China, Hefei 230088, China}

\author{Jian-Wei Pan}%
\affiliation{Hefei National Research Center for Physical Sciences at the Microscale and School of Physical Sciences, University of Science and Technology of China, Hefei 230026, China}
\affiliation{CAS Center for Excellence in Quantum Information and Quantum Physics, University of Science and Technology of China, Hefei 230026, China}
\affiliation{Hefei National Laboratory, University of Science and Technology of China, Hefei 230088, China}

\date{\today}

\begin{abstract}

It is challenging to build scalable quantum processors capable of both parallel control and local operation.
As a promising platform to overcome this challenge, optical lattices offer exceptional parallelism. However, it has been struggling with precise local operations due to relatively narrow lattice spacings.
Here, we introduce a new quantum processor incorporating orbit-qubit encoding and internal states (as auxiliary degrees of freedom) to achieve spatially selective operations together with parallel control.
With this processor, we generate one-dimensional and two-dimensional cluster states using minimal layers of controlled-Z gates.
We experimentally detect the multipartite entanglement of a two-dimensional cluster state involving 123 orbit qubits through direct stabilizer measurements, verifying the full bipartite non-separability.
Furthermore, we demonstrate measurement-based quantum computation by implementing single-qubit and two-qubit logical gates, highlighting the flexibility of orbit-qubit operations. 
Our results establish orbit-qubit optical lattices as a scalable quantum processing architecture, opening new pathways for quantum computation applications.

\end{abstract}

\maketitle
\droptocpage


\section{\label{sec:intro}Introduction}

Quantum computers have the potential to solve practical problems that are intractable for their classical counterparts \cite{nielsen2010quantum,daley2022practical}. 
The past decades have witnessed substantial experimental advancements in quantum computation with various physical platforms, including trapped ions \cite{moses2023race,guo2024site,paetznick2024demonstration}, superconducting quantum circuits \cite{arute2019quantum,cao2023generation}, photonic systems \cite{zhong2020quantum,madsen2022quantum}, and neutral atoms \cite{bluvstein2022quantum,evered2023high,bluvstein2024logical}.
Ultracold neutral atoms trapped in optical lattices \cite{jaksch1998cold,greiner2002quantum,duan2003controlling,anderlini2007controlled,trotzky2008time,dai2016generation} represent a promising platform owing to their excellent scalability and exceptionally long coherence times \cite{yang2020cooling,bojovic2025high,kiefer2025protected}.
Recent advancements in quantum gas microscopy have enabled site-resolved addressing and detection, thus facilitating the direct verification of multipartite entanglement generated by high-fidelity $\sqrt{\mathrm{SWAP}}^{\dag}$ gates \cite{zhang2023scalable}. 
However, further performing computational tasks is still elusive. 

In the context of measurement-based quantum computation, a universal computing task is realized by preparing cluster states and then implementing designed single-qubit measurements and relevant rotations \cite{raussendorf2001one,raussendorf2003measurement,raussendorf2007topological,walther2005experimental}.
As the typical lattice spacing between neighboring sites approaches the optical diffraction limits of current quantum gas microscopy, this inevitably causes operational crosstalk during attempts of single-site manipulation. 
Consequently, the inadequate fidelity of local operations poses a significant barrier to accurately verify cluster states, as well as restricting their further computational applications.

\begin{figure*}[t!]
  \centering
  \includegraphics[width=1\textwidth]{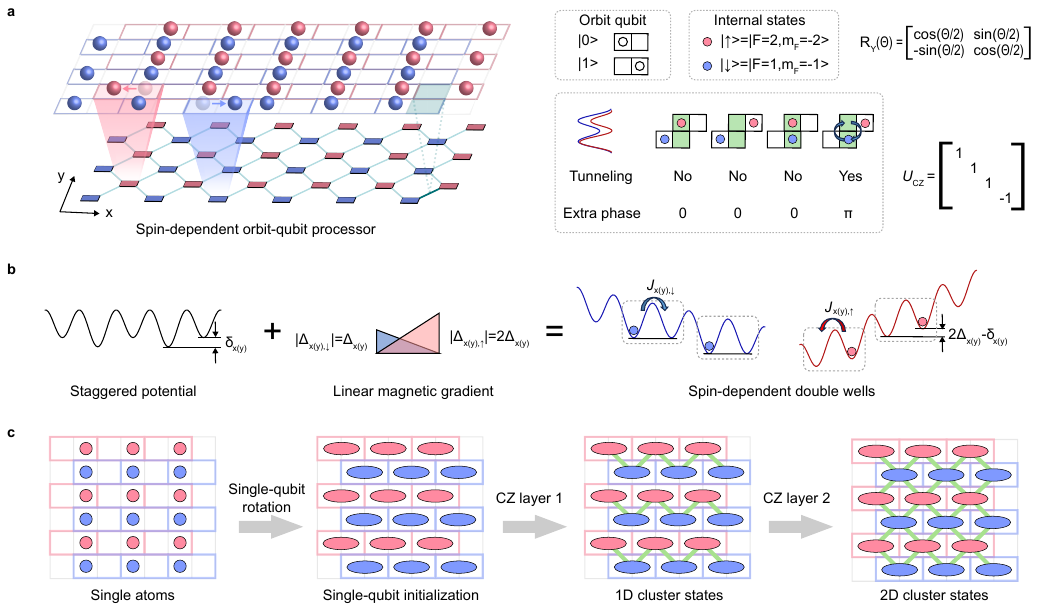}
  \caption{
  \textbf{Spin-dependent orbit-qubit quantum processor in optical lattices}.
  \textbf{a}, Schematic diagram of a spin-resolved orbit-qubit quantum processor. 
  Each group of connected squares represents an effective double-well potential encoding an orbit qubit, whose computational basis states, $\ket{0}$ and $\ket{1}$, correspond to the atom occupying the left or right well, respectively. 
  The orbit qubits are implemented via coherent single-atom tunneling of a $^{87}\mathrm{Rb}$ atom in internal hyperfine states $\ket{\downarrow}=\ket{F=1,m_F=-1}$ (blue circles) or $\ket{\uparrow}=\ket{F=2,m_F=-2}$ (red circles) along the $x$-direction. 
  Green shading highlights an effective double-well potential along the $y$-direction for implementing the controlled-Z (CZ) gates. 
  Under selective resonant tunneling of $\ket{\downarrow}$ atoms, based on the conditional occupancy of the adjacent sites, a conditional phase of $\pi$ accumulates after one complete tunneling cycle, thus realizing an effective CZ gate between the adjacent orbit qubits.   
  \textbf{b}, Spin-dependent effective double-well potential realized by combining staggered potentials ($\delta$, with $\delta \gg J$) and magnetic gradient. 
  Resonant tunneling is enabled when $\delta$ approximately matches the magnetic energy difference between adjacent sites, $2\Delta$ for $\ket{\uparrow}$ atoms or $\Delta$ for $\ket{\downarrow}$ atoms (depicted in \textbf{b} right), thus forming effective double wells (gray dashed rectangles).
  \textbf{c}, Experimental roadmap for generating cluster states.
  Atoms are first arranged in alternating columns along the $x$-direction and initialized in a N\'eel-type spin configuration, encoding two types of orbit qubits. 
  Global $R_Y(\pi/2)$ gates sequentially applied to orbit qubits with different spin states prepare them in the superposition of $(\ket{0} + \ket{1}) / \sqrt{2}$. 
  A single layer of CZ gates along the $y$-direction yields copies of 1D cluster states, while two successive layers create a 2D cluster state. 
  }
  \label{fig1:Encoding}
\end{figure*}

In this work, we demonstrate a programmable quantum processor based on spin-resolved orbit qubits in optical lattices (Fig.~\ref{fig1:Encoding}).  
The orbit qubits \cite{impertro2024local} are encoded by the spatial occupation of individual $^{87}$Rb atoms in spin-dependent double-well potentials, created through a combination of staggered optical potentials and linear magnetic field gradients. 
The internal hyperfine state of each $^{87}$Rb atom serves as an auxiliary degree of freedom, enabling selective addressing and tailored initialization for specific computational tasks. 
The flexible initialization of both orbit and spin configurations facilitates efficient quantum circuits and enhanced local manipulations. 
Leveraging these capabilities, we generate one-dimensional (1D) and two-dimensional (2D) cluster states across a region exceeding $10\times 20$ sites, using only one or two layers of controlled-Z (CZ) gates. 
We verify the created entanglement by measuring the corresponding three- and five-body stabilizers. 
Furthermore, we demonstrate measurement-based quantum computation on these cluster states, implementing both single-qubit and two-qubit logical gates. 
Our results demonstrate the flexibility of the orbit-qubit quantum processor, establishing a new framework towards scalable quantum computation.

\section{\label{sec:encode}Orbit qubits in spin-dependent double-well potentials}

\subsection{Double wells generated by a staggered potential and a linear magnetic gradient}

We encode orbit qubits using a double-well potential created by combining optical superlattices with a linear gradient tilt (Fig.~\ref{fig1:Encoding}b). 
The optical superlattice is produced by superimposing short- and long-wavelength lattices, with a wavelength ratio of 2 (details are provided in our previous work \cite{zhang2023scalable} and Methods). 
In this work, we operate in a staggered configuration by aligning the minima of both lattices, such that the potential difference between adjacent lattice sites is solely controlled by the depth of the long lattice. 
An additional magnetic gradient induces a linear tilt across the lattice. 
In this case, the single-particle Hamiltonian of an atom with spin $\sigma$ can be written as
\begin{align}
\hat{H}_{\mathrm{EDW}} =& -J\sum_{j=0}^{N-1}{\left(\hat{a}^{\dag}_{j,\sigma}\hat{a}_{j+1,\sigma}+h.c.\right)}\notag \\
&+ \frac{\delta}{2}\sum_{j=0}^{N-1}{(-1)^j \hat{n}_{j,\sigma}}+\sum_{j=0}^{N-1}{j\Delta_{\sigma}\hat{n}_{j,\sigma}},
\end{align}
\noindent where $\hat{a}^{(\dag)}_{j,\sigma}$ is the annihilation (creation) operator and $\hat{n}_{j,\sigma}$ is the atom number operator for site $j$ and spin $\sigma$. 
$J$ denotes the tunneling amplitude, $\delta$ represents the staggered potential, and $\Delta_{\sigma}$ corresponds to the magnetic gradient potential. 
When $|\Delta_{\sigma}| \gg J$ and $\delta \approx |\Delta_{\sigma}|$, the potential becomes isolated double wells.
Therefore, the orbit qubit can be defined by an atom sitting in the left or right wells of the double-well potentials, as shown in Fig.~\ref{fig1:Encoding}a, with computational basis states $\ket{0}$ and $\ket{1}$ corresponding to the particle occupying the left or right site, respectively.

In our experiment, we start by preparing copies of nearly defect-free arrays of ultracold bosonic $^{87}$Rb atoms, arranged in alternating columns along the $x$-direction. 
The atoms are then initialized in arbitrary spin configurations drawn from two hyperfine ground states, $\ket{\uparrow} = \ket{F=2, m_F=-2}$ and $\ket{\downarrow} = \ket{F=1, m_F=-1}$, using spin-dependent superlattices in combination with site-resolved addressing. 
To realize spin-dependent control, we apply a magnetic field gradient oriented diagonally in the $x$-$y$ plane. 
Owing to the differences in Land\'e $g$-factors and magnetic moments of the two spin states, the resulting energy offsets between adjacent sites satisfy $\Delta_{\uparrow} = -2\Delta_{\downarrow}$. 
This configuration yields tunable energy gradients along both $x$- and $y$-lattice directions: for $\ket{\downarrow}$ atoms, the site-to-site energy difference is $\Delta_{x(y)}$, while for $\ket{\uparrow}$ atoms, it is $2\Delta_{x(y)}$ along the $x$($y$)-direction and in opposite directions.

By adjusting the depth of the long lattice potential, we engineer spin-selective double-well configurations. 
As shown in Fig.~\ref{fig1:Encoding}b, along the $x$-direction, setting $\delta_x = \Delta_x$ enables resonant tunneling for $\ket{\downarrow}$ atoms between neighboring sites to their right, forming a balanced double well. 
In this regime, $\ket{\uparrow}$ atoms are off-resonant and remain localized. 
Conversely, when $\delta_x = 2\Delta_x$, tunneling becomes resonant for $\ket{\uparrow}$ atoms to their left neighbor, while $\ket{\downarrow}$ atoms remain localized. 
An analogous scheme can also be implemented along the $y$-direction. 
This configuration allows for independent and selective encoding of orbit qubits in both spin species, enabling the preparation of two-dimensional cluster states by sequentially applying layers of controlled-Z (CZ) gates along the $y$-axis (see below). 
The brick-wall connectivity minimizes the circuit depth required for entanglement generation and is naturally suited for stabilizer measurements based on spin-resolved operations.

\begin{figure}[h!]
  \centering
  \includegraphics[width=0.45\textwidth]{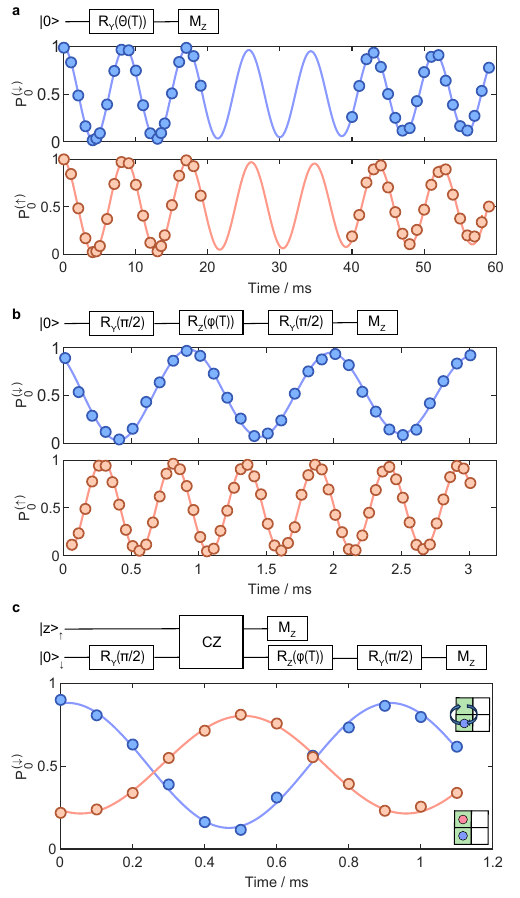}
  \caption{\textbf{Quantum gate operations on orbit qubits.}
  \textbf{a}, Single-qubit $R_Y$ rotation realized by resonant tunneling in the corresponding balanced double-well potentials for two types of orbit qubits. 
  The population remaining in the initial site oscillates with tunneling amplitudes of $J_{x,\downarrow}=58.00(5)~\mathrm{Hz}$ and $J_{x,\uparrow}=56.91(6)~\mathrm{Hz}$ for $\ket{\downarrow}$ (top panel) and $\ket{\uparrow}$ (bottom panel), respectively; fitted decay times are $265(25)~\mathrm{ms}$ and $225(26)~\mathrm{ms}$. 
  \textbf{b}, Single qubit $R_Z$ rotation implemented by phase accumulation in a tilted double well potential for two types of orbit qubits. 
  Measured Ramsey fringers for the two spin states oscillate at $954(3)~\mathrm{Hz}$ (top panel) and $1893(2)~\mathrm{Hz}$ (bottom panel), respectively, consistent with the expected ratio $|\Delta_{x,\uparrow}| = 2|\Delta_{x,\downarrow}|$. 
  \textbf{c}, Controlled-Z gates are evaluated by conditional tunneling along the $y$-direction. 
  For $\ket{\downarrow}$ atoms, almost opposite-phased oscillation behavior is observed when the neighbouring site is empty (blue line) compared with occupied (orange line), as marked in the inset. 
  Such a $\pi$ phase shift evidences the implementation of an effective CZ gate between adjacent orbit qubits.}
  \label{fig2:Gate}
\end{figure}

\subsection{Single-qubit gates via coherent orbit dynamics}

Using the above encoding protocol, we realize two kinds of orbit qubits along the $x$-direction, each associated with one of the two spin states. 
To evaluate their performance, we first characterize single-qubit gate operations \cite{impertro2024local}. 
The rotation around the $y$-axis of the Bloch sphere, $R_Y$, corresponds to coherent single-atom tunneling within the double wells. 
In contrast, the $R_Z$ rotations around the $z$-axis are realized by inducing a controllable energy shift between the left and right orbitals. 
This results in a relative phase accumulation during free evolution in a tilted potential.

To implement and characterize the $R_Y$ gate, we initialize atoms in either the $\ket{\downarrow}$ or $\ket{\uparrow}$ state and ramp up the magnetic gradient to introduce the site-to-site energy offset to $\Delta_x$ (for $\ket{\downarrow}$) along the $x$-direction. 
We then adjust the long lattice depth to $\delta_x = \Delta_x$ for $\ket{\downarrow}$ atoms and $\delta_x = 2\Delta_x$ for $\ket{\uparrow}$ atoms, ensuring a balanced double-well configuration for both spin states. 
The tunneling dynamics are activated by quenching the depth of the short lattice. 
After an evolution time, the dynamics are frozen, and the final atomic distributions are read out via site-resolved imaging using a quantum gas microscope.
To mitigate the initial state imperfections and atom loss, we post-select configurations with single-atom occupation in the double wells (the post-selection rule is also applied in the subsequent measurements of correlations).
Fig.~\ref{fig2:Gate}a shows the observed tunneling oscillations with amplitudes $J_{x,\downarrow} = 58.00(5)~\mathrm{Hz}$ and $J_{x,\uparrow} = 56.91(6)~\mathrm{Hz}$, satisfying the localization condition $J_x \ll \delta_x, \Delta_x$. 
In a region of $14 \times 14$ lattice sites, the oscillation decay times are extracted to be $265(25)~\mathrm{ms}$ for $\ket{\downarrow}$ and $225(26)~\mathrm{ms}$ for $\ket{\uparrow}$, indicating long coherence and enabling single-qubit $\hat{R}_Y(\pi/2)$ gates with an average fidelity of $99.11(7)\%$.

To implement $R_Z$ gates, we use a Ramsey-type sequence to induce and characterize the phase accumulation between the orbit states. 
After applying a $\hat{R}_Y(\pi/2)$ gate to prepare a superposition of $\ket{0}$ and $\ket{1}$ for each kind of orbit qubit, we suppress tunneling by suddenly increasing the short lattice depth and quenching the staggered potential to $\delta_x = 0$. 
In this tilted configuration, the orbit qubit precesses around the $z$-axis of the Bloch sphere at a frequency determined by the energy offset $\Delta_x$ for $\ket{\downarrow}$ and $2\Delta_x$ for $\ket{\uparrow}$. 
After a variable evolution time $T$, a second $\hat{R}_Y(\pi/2)$ gate is applied, followed by site-resolved measurement of the final atomic distribution. 
As shown in Fig.~\ref{fig2:Gate}b, the Ramsey fringes for $\ket{\downarrow}$ and $\ket{\uparrow}$ exhibit oscillation frequencies of $954(3)~\mathrm{Hz}$ and $1893(2)~\mathrm{Hz}$, respectively, consistent with the expected relation $|\Delta_{x,\uparrow}| = 2|\Delta_{x,\downarrow}|$. 
These results confirm the controlled phase accumulation essential for high-fidelity $R_Z$ gate operations.

\subsection{CZ gates based on controlled single-atom tunneling}

After establishing high-fidelity single-qubit operations, we now demonstrate the implementation of two-qubit gates. 
Our scheme leverages controlled single-atom tunneling between neighboring orbit qubits to realize a controlled-Z (CZ) gate, a fundamental building block for generating large-scale entangled states. 
As shown in Fig.~\ref{fig1:Encoding}a, in our encoding architecture, two-qubit entangling gates are implemented along the $y$-direction, coupling adjacent orbit qubits encoded in different spin states. 
When the staggered potential is set to $\delta_y = \Delta_y$, $\ket{\downarrow}$ atoms become resonant for coherent tunneling between double wells, but only when the adjacent lattice site is unoccupied; otherwise, the on-site interaction $U~(U=447(2)~\mathrm{Hz})$ between these two spin states significantly suppresses the tunneling process. 
In contrast, $\ket{\uparrow}$ atoms remain off-resonant due to a larger detuning. 
As a result, tunneling occurs only for one specific input configuration, leading to a conditional dynamical phase. 
A complete tunneling cycle (a 2$\pi$-pulse) accumulates an additional phase of $\pi$, realizing an effective CZ gate between the two orbit qubits.
Note that such a process can also introduce additional $\hat{Z}$ gates for specific qubits. For simplicity, we directly regard such operations as CZ gates and reinterpret the measurement results according to the additional $\hat{Z}$ gates.

\begin{figure*}[t!]
  \centering
  \includegraphics[width=1\textwidth]{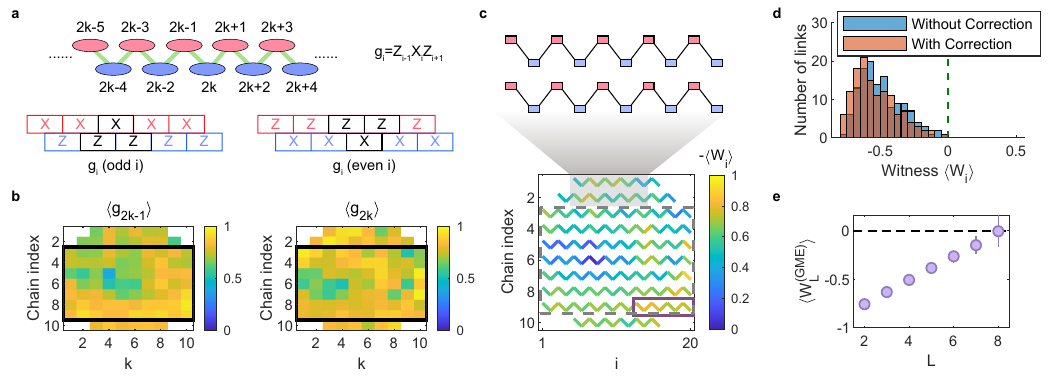}
  \caption{
  \textbf{Generation and verification of 1D cluster states}.
  \textbf{a}, Schematic of a 1D cluster state (top) and the two local measurement settings used to access the three-body stabilizer $\hat{g}_i$ (bottom). 
  Red and blue rectangles indicate selective rotations applied to $\ket{\uparrow}$ and $\ket{\downarrow}$ orbit qubits, respectively, before site-resolved readout. 
  \textbf{b}, Spatial map of stabilizer values.
  Within the region outlined in the black rectangles, we obtain averages of $0.769(1)$ for odd $i$ and $0.775(1)$ for even $i$. 
  \textbf{c}, Map of entanglement witness $\langle\hat{W}_i\rangle$. 
  Links are marked in color where $\langle\hat{W}_i\rangle < 0$, indicating entanglement between adjacent orbit qubits. 
  Grey and purple rectangles mark data sets analyzed in \textbf{d} and \textbf{e}, respectively. 
  \textbf{d}, Histogram of witness $\langle\hat{W}_i\rangle$ in grey region of \textbf{c}; all values are negative, confirming \textit{full bipartite non-separability} across the entire chain of at least 20 qubits. 
  \textbf{e}, Genuine multipartite entanglement (GME) witness $\langle \hat{W}^{(\mathrm{GME})}_{L} \rangle$ versus chain length $L$. 
  As values remain negative up to $L=7$, verifying at least $7$-qubit GME. 
  The brown dashed line marks the threshold for detecting GME. 
  }
  \label{fig3:Cluster}
\end{figure*}

To verify this conditional phase accumulation, we perform the quantum circuit in Fig.~\ref{fig2:Gate}c under different initial spin configurations. 
Starting from a N\'eel-type spin configuration along the $y$-direction, we perform two kinds of experiments: in the first one, we retain both the spin states; and in the other, we selectively remove the $\ket{\uparrow}$ atoms by using a resonant laser pulse. 
We then set $\delta_y = \Delta_y$ and quench the short lattice potential to activate the single-atom tunneling dynamics for $\ket{\downarrow}$ atoms along the $y$-direction. 
After holding for a controlled evolution time, corresponding to a 2$\pi$-pulse, we freeze the tunneling dynamics and apply $\hat{R}_Y(\pi/2)\hat{R}_{Z}(\varphi)$ to the $\downarrow$ qubit.
The final atom distributions are measured via site-resolved imaging. 
Fig.~\ref{fig2:Gate}c shows the resulting oscillations for the two configurations. 
A clear opposite-phase behavior is observed, confirming the presence of a relative $\pi$ phase shift induced by the conditional tunneling.
In Sec.~\ref{sec:cluster}a, we measure the stabilizers of 1D cluster states, and the average fidelity of the CZ gates is $94.8(13)\%$ in the purple rectangle region in Fig.~\ref{fig3:Cluster}c (See Methods).

Moreover, by tuning the detuning offset $\delta_y - \Delta_y$ and the duration of the tunneling process, the accumulated phase can be continuously controlled, enabling the implementation of general controlled-phase (CPhase) gates. 
Notably, this controlled collisional CPhase gate scheme is not limited to orbit qubits and may also be extended to spin-qubit-based quantum processors. 
Compared to conventional two-qubit entangling gates based on second-order spin exchange dynamics implemented in spin-qubit-based optical lattice platforms, our orbit-qubit encoding methods operate on a timescale determined by the bare tunneling rate of $1/J$, offering a path toward faster and more robust gates against inhomogeneities, and thus potentially achieving higher gate fidelity.

\section{Preparing 1D and 2D cluster states in optical lattices}
\label{sec:cluster}

Having established the essential components, encoding orbit qubits, performing high-fidelity single-qubit rotations, and realizing CZ entangling gates, we now turn to the preparation and verification of cluster states, a fundamental resource for measurement-based quantum computation (MBQC) \cite{raussendorf2001one,raussendorf2003measurement,raussendorf2007topological}. 

Fig.~\ref{fig1:Encoding}c shows the procedure for generating 1D and 2D cluster states using our brick-wall qubit architecture. 
The experiment begins by loading $^{87}\mathrm{Rb}$ atoms into alternating columns along the $x$-direction. 
A spin-dependent superlattice potential is then used to initialize a N\'eel-type spin configuration along the $y$-direction (see Methods). 
A magnetic gradient is applied diagonally across the $x$–$y$ plane, resulting in spin-dependent energy tilts that, together with tailored long lattice depths, form double wells arranged in a brick-wall pattern for the two spin states. 

To initialize each orbit qubit in the superposition state $\ket{+} = (\ket{0} + \ket{1})/\sqrt{2}$, we sequentially apply global $\hat{R}_Y(\pi/2)$ gates to both spin species. 
Entanglement is subsequently generated by applying two layers of CZ gates along the $y$-direction. 
In the first layer, we set $\delta_y = \Delta_y$, enabling $\ket{\downarrow}$ atoms to undergo a 2$\pi$ tunneling cycle in their respective double wells, thereby entangling adjacent qubits into copies of 1D cluster states. 
In the second layer, we adjust the long lattice depth to $\delta_y = 2\Delta_y$ to activate the tunneling of $\ket{\uparrow}$ atoms, thus completing the construction of the 2D cluster state.

\subsection{1D cluster states}

As mentioned above, the brick-wall geometry of our orbit qubit array enables efficient preparation of 1D cluster states through only a single layer of CZ gates applied along the $y$-direction. 
To verify the resulting entangled states, we measure the three-body stabilizers $\hat{g}_{i} = \hat{Z}_{i-1}\hat{X}_i\hat{Z}_{i+1}$, which characterize the standard 1D cluster state. 
These observables are accessed using only two complementary local measurement settings, as shown in Fig.~\ref{fig3:Cluster}a, implemented through spin-selective operations. 
In the first setting (Fig.~\ref{fig3:Cluster}a bottom left), we set $\delta_x = 2\Delta_x$ to activate a global $\hat{R}_Y(\pi/2)$ gate over the $\ket{\uparrow}$ atoms, while the $\ket{\downarrow}$ atoms remain unperturbed. 
This operation rotates the orbit qubits defined by the $\ket{\uparrow}$ atoms for the $X$ measurement, while preserving the $Z$ measurement for those orbit qubits defined by $\ket{\downarrow}$ atoms. 
Therefore, this enables the evaluation of $\langle \hat{g}_i \rangle$ for odd $i$. 
In the second setting (bottom right), the pulse is instead applied to the $\ket{\downarrow}$ atoms, enabling access to stabilizers for even $i$.
To mitigate the error of $X$ measurement, we locally correct the values of the stabilizer measurement (See Methods). 

Fig.~\ref{fig3:Cluster}b shows the measured results of $\langle \hat{g}_i \rangle$ across the whole atom arrays. 
Within the region marked by black rectangles, we identify seven independent 20-qubit chains. 
The average values of the stabilizer $\langle \hat{g}_i \rangle$ for odd and even $i$ are $0.769(1)$ and $0.775(1)$, respectively. 
To verify the entanglement between neighboring orbit qubits $\{i,i+1\}$, we evaluate the following entanglement witness 
\begin{align}
\hat{W}_i = \hat{\mathcal{I}} - \hat{Z}_{i-1}\hat{X_i}\hat{Z}_{i+1} - \hat{Z}_{i}\hat{X}_{i+1}\hat{Z}_{i+2}.
\label{EW1D}
\end{align}
If $\langle \hat{W}_i \rangle < 0$ holds for any $i$, the state is a multipartite entangled state, also known as \textit{full bipartite nonseparability} \cite{zhou2022scheme}.

As shown in Fig.~\ref{fig3:Cluster}c, we plot the spatial distribution of $\langle \hat{W}_i \rangle$ across the whole atom array, where the color links mark the measured values. 
And in Fig.~\ref{fig3:Cluster}d, we show the statistical distribution of $\langle \hat{W}_i \rangle$ in the region highlighted by the gray dashed rectangle. 
All evaluated witness values are negative, which verifies the presence of multipartite entanglement in 1D chains of at least 20 orbit qubits. 
The spatial variation in $\langle \hat{g}_i \rangle$ primarily originates from residual inhomogeneity in the local chemical potential (see Methods).

To further detect the presence of \textit{genuine multipartite entanglement} (GME), we use the following GME witness
\begin{align}
\hat{W}^{(\mathrm{GME})}_{L} = (L-1)\hat{\mathcal{I}} - \sum_{s=1}^{L}\hat{g}_s.
\label{EWGME0}
\end{align}
Fig.~\ref{fig3:Cluster}e presents the average values of $\hat{W}^{(\mathrm{GME})}_{L}$ as a function of the segment length $L$ within the region marked by the purple rectangle in Fig.~\ref{fig3:Cluster}c. 
We observe clear signatures that exceed the threshold for segments with up to 7 orbit qubits, confirming the existence of seven-partite GME.

\subsection{2D cluster states}

\begin{figure}[t!]
  \centering
  \includegraphics[width=0.45\textwidth]{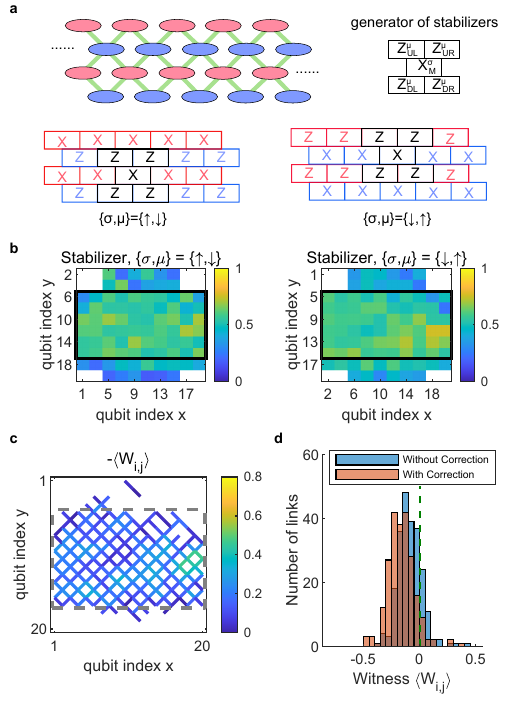}
  \caption{
  \textbf{Generation and detection of 2D cluster states}. 
  \textbf{a}, Schematic of a 2D cluster state (top) and the two local measurement settings used to access the five-body stabilizer (bottom). 
  Red and blue rectangles represent selective rotations applied to $\ket{\uparrow}$ and $\ket{\downarrow}$ orbit qubits, respectively, before site-resolved detection. 
  The inset defines the stabilizer centered on site $M$ (spin $\sigma$) with four neighbors of spin $\mu$: $\hat{Z}^{\mu}_{DL}\hat{Z}^{\mu}_{DR}\hat{X}^{\sigma}_{M}\hat{Z}^{\mu}_{UL}\hat{Z}^{\mu}_{UR}$, where $DL$, $DR$, $UL$, and $UR$ label the lower-left, lower-right, upper-left, and upper-right neighbors, respectively, and $\sigma, \mu \in \{\uparrow,\downarrow\}$ denote the spin states. 
  \textbf{b}, Spatial maps of the five-body stabilizer for the $\{\uparrow,\downarrow\}$ (left) and $\{\downarrow,\uparrow\}$ (right) configurations.
  \textbf{c}, Map of entanglement witness $\langle \hat{W}_{i,j} \rangle$ for the 2D state. 
  Links are marked in color where $\langle \hat{W}_{i,j} \rangle < 0$ within the black rectangle, signifying a
  connected entangled region spanning 123 orbit qubits (exceeding the threshold by one standard deviation) and demonstrating large-scale \textit{full bipartite nonseparability}.
  \textbf{d}, Histogram of witness $\langle \hat{W}_{i,j} \rangle$ from the marked region in \textbf{c}, where most values are negative supports the large-scale 2D entanglement. 
  }
  \label{fig4:Cluster}
\end{figure}

As mentioned before, building upon the 1D construction, 2D cluster states are generated by applying an additional layer of CZ gates (let $\delta_y = 2\Delta_y$) that couple neighboring orbit qubits between adjacent 1D chains. 
The resulting entangled state can be verified using the same two complementary local measurement settings employed in the 1D case. 
However, the relevant stabilizers now involve five-body operators centered on each orbit qubit, as illustrated in Fig.~\ref{fig4:Cluster}a. 
The stabilizer for a central site $M$ with spin $\sigma$ and four neighboring sites with spin $\mu$ takes the form $\hat{Z}^{\mu}_{DL}\hat{Z}^{\mu}_{DR}\hat{X}^{\sigma}_{M}\hat{Z}^{\mu}_{UL}\hat{Z}^{\mu}_{UR}$. 
Here, $DL$, $DR$, $UL$, and $UR$ label the lower-left, lower-right, upper-left, and upper-right neighbors, respectively, and $\sigma, \mu \in \{\uparrow,\downarrow\}$ denote the spin states. 
Fig.~\ref{fig4:Cluster}b shows the spatial map of stabilizer values measured under two spin configurations, $\{\uparrow,\downarrow\}$ and $\{\downarrow,\uparrow\}$. 
In the regions marked by black rectangles, the average stabilizer values are $0.572(7)$ and $0.584(6)$, respectively.

To further verify the multipartite entanglement across the 2D array, similarly to the 1D case, we use the following witness 
\begin{align}
\hat{W}_{i,j} = \hat{\mathcal{I}} - \hat{X}_i\hat{Z}_j\bigotimes_{\left\langle i,s\right\rangle,s\neq j}\hat{Z}_{s}  - \hat{Z}_i\hat{X}_j\bigotimes_{\left\langle j,s\right\rangle,s\neq i}\hat{Z}_{s}.
\label{EW2D}
\end{align} 
This witness detects bipartite entanglement between adjacent qubits $i$ and $j$. 
Fig.~\ref{fig4:Cluster}c presents the corresponding spatial distribution of $\langle \hat{W}_{i,j} \rangle$ across the 2D arrays.
The histogram in Fig.~\ref{fig4:Cluster}d shows the statistical distribution of witness values within the gray-marked region. 
We clearly observe widespread negative values of $\langle \hat{W}_{i,j} \rangle$ across a 2D region containing up to 123 orbit qubits, thus verifying the multipartite entanglement over a large-scale 2D system (details of 2D entanglement detection refer to Methods).

\section{Demonstration of one-way quantum computation}

\begin{figure*}[t!]
  \centering 
  \includegraphics[width=1\textwidth]{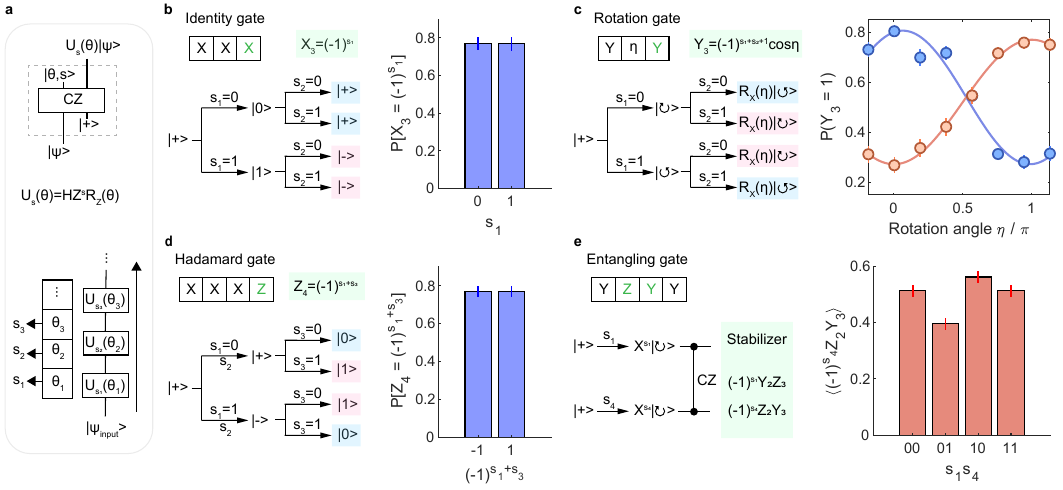}
  \caption{
  \textbf{Measurement-based quantum logic operation on orbit-qubit cluster states}.
  \textbf{a}, Schematic of quantum information propagation based on measurement. 
  Top: A CZ gate entangles an arbitrary state $\ket{\psi}$ with an ancilla $\ket{+}$; subsequent projective measurement of the first qubit teleports the state to the second. 
  Bottom: Repeating such measurements propagates information along a cluster state.
  \textbf{b}, Logical identity gate on a three-qubit cluster measured in $X_1 \otimes X_2 \otimes X_3$ basis. 
  Bars indicate the probability of obtaining the correct $X_3$ outcome for the two possible measured results, $s_1=0$ and $s_1=1$. 
  The average successful probability is 0.771(26). 
  \textbf{c}, Logical single-qubit rotation gate implemented under the basis $Y_1\otimes (\mathcal{B_{\eta}})_2 \otimes Y_3$.
  Measured $Y_3$ oscillates follow the expected sinusoid curve, indicating a coherent rotation operation. 
  \textbf{d}, Logical Hadamard gate using a four-qubit cluster measured in $X_1 \otimes X_2 \otimes X_3 \otimes Z_4$. 
  Bars show the output probability of $Z_4$ conditioned on the measured results from the first and third qubits, $s_1$ and $s_3$. 
  \textbf{e}, Logical entangling gate from a four-qubit cluster measured in $Y_1\otimes Z_2\otimes Y_3\otimes Y_4$. 
  Qubits 1 and 4 serve as logical inputs, while qubits 2 and 3 form the entangled output pair. 
  Bars show the stabilizer $\hat{Z}_2\hat{Y}_3$ based on the measured results from qubits 1 and 4, $s_1$ and $s_4$, verifying correlation between qubits 2 and 3. 
  }
  \label{fig5:MBQC}
\end{figure*}

Following the successful generation of large-scale cluster states in our spin-dependent orbit-qubit processor, we now present an implementation of measurement-based quantum computation (MBQC). 
Although experimental imperfections currently limit the execution of full-scale quantum algorithms, our system provides an ideal platform for exploring fundamental logical gate operations based on measurement-driven quantum information processing.

Before delving into the details of experimental implementations, we provide a brief explanation of the principle of logical gate operations within MBQC frameworks.  
As shown in Fig.~\ref{fig5:MBQC}a, after entangling the left arbitrary input state $\ket{\psi}$ and the right ancillary qubit initialized as $\ket{+}$ with a CZ gate, measuring the input qubit results in an effective operation on the ancillary qubit.
Assume the input qubit is measured in the basis of
\begin{align*}
\mathcal{B_{\theta}} = \left\{\frac{\ket{0}+e^{i\theta}\ket{1}}{\sqrt{2}},\frac{\ket{0}-e^{i\theta}\ket{1}}{\sqrt{2}}\right\},
\end{align*}
and the measurement outcome is $s$.
The input qubit collapses to
\begin{align*}
\ket{\theta,s} = \frac{\ket{0}+(-1)^s e^{i\theta}\ket{1}}{\sqrt{2}}.
\end{align*}
\noindent Then the ancillary qubit is transformed by the operation (see Methods)
\begin{align*}
\hat{U}_s(\theta) = \hat{H}\hat{Z}^s\hat{R}_Z(\theta).
\end{align*}
In this manner, single-qubit measurements in rotated bases propagate quantum information along a 1D cluster state. 
In the following, we let $s_i$ denote the measurement result of qubit $i$. 
Cluster states of desired length are deterministically prepared using site-selective addressing techniques (see Methods). 
The protocols for implementing the following measurement settings are described in the Methods section.

We first experimentally demonstrate an identity gate using a 3-qubit cluster state measured in the $X_1 \otimes X_2 \otimes X_3$ basis. 
In our implementation, the logical input state is initialized as $\ket{\psi_{\mathrm{input}}} = \ket{+}$ on the first qubit, with the third qubit serving as the logical output.
As shown in Fig.~\ref{fig5:MBQC}b, the output state is $\ket{+}$ for $s_1=0$ and $\ket{-}$ for $s_1=1$.
Bars in Fig.~\ref{fig5:MBQC}b show the probabilities of correct $X_3$ measurement for different measurement results of $s_1$. 
The average correct probability is $0.771(26)$.

Then, we demonstrate single-qubit rotation gates based on $3$-qubit cluster states under the measurement basis $Y_1\otimes (\mathcal{B_{\eta}})_2 \otimes Y_3$ (Fig.~\ref{fig5:MBQC}c).
The output state is $\hat{R}_X(\eta)\ket{\circlearrowleft}$ for $s_1+s_2=0$ and $\hat{R}_X(\eta)\ket{\circlearrowright}$ for $s_1+s_2=1$ ($\ket{\circlearrowright/\circlearrowleft}=(\ket{0}\pm i\ket{1})/\sqrt{2}$, and the addition is modulo $2$), leading to $\langle \hat{Y_3} \rangle = (-1)^{s_1+s_2+1}\cos{\eta}$.
As shown in Fig.~\ref{fig5:MBQC}c, $Y_3$ oscillates following the expected sinusoid curve.

We further demonstrate logical Hadamard gates based on $4$-qubit cluster states under the measurement basis $X_1 \otimes X_2 \otimes X_3 \otimes Z_4$ (Fig.~\ref{fig5:MBQC}d). 
In this setting, the output state is $\ket{0}$ for $s_1+s_3=0$ and $\ket{1}$ for $s_1+s_3=1$.
Fig.~\ref{fig5:MBQC}d shows the probabilities of correct $Z_4$ measurement for different measurement results of $s_1+s_3$. The average correct probability is $0.768(20)$.

Finally, we demonstrate two-qubit entangling gates based on $4$-qubit cluster states under the measurement basis $Y_1\otimes Z_2\otimes Y_3\otimes Y_4$. 
Here, the first and fourth qubits correspond to the logical input qubits and $\ket{\Psi_{\mathrm{input}}} = \ket{+}\ket{+}$. The output states (qubits $\{2,3\}$) are
\begin{align*}
\ket{\Psi_{\mathrm{output}}} = \hat{U}_{CZ}\left[\frac{\ket{0}+i(-1)^{s_1}\ket{1}}{\sqrt{2}}\otimes \frac{\ket{0}+i(-1)^{s_4}\ket{1}}{\sqrt{2}}\right],
\end{align*}
with stabilizers $(-1)^{s_1}\hat{Y}_2\hat{Z}_3$ and $(-1)^{s_4}\hat{Z}_2\hat{Y}_3$.
Fig.~\ref{fig5:MBQC}e shows measured $\langle \hat{Z}_2\hat{Y}_3 \rangle$ for different $\{s_1,s_4\}$. It is clear that the output qubits are correlated and the sign of $\langle \hat{Z}_2\hat{Y}_3 \rangle$ is dependent on $s_4$. The average value of $\langle (-1)^{s_4}\hat{Z}_2\hat{Y}_3 \rangle$ is $0.495(11)$.

\section{Summary and Discussion}

In summary, we demonstrated a programmable quantum processor based on spin-resolved orbit qubits in optical lattices. 
We realized coherent single-qubit operations, high-fidelity CZ entangling gates, and prepared entangled 1D and 2D cluster states, confirming multipartite entanglement involving up to 123 qubits. 
Moreover, we implemented essential logical quantum gates, including identity, rotations, Hadamard, and two-qubit entangling operations, via local measurements on 1D cluster state segments, highlighting the potential of orbit-qubit architectures for scalable MBQC.

In the future, further improvements in gate operation fidelity, combined with mid-circuit measurements \cite{deist2022mid,graham2303mid,bluvstein2024logical}, will push forward practical one-way quantum computation. 
Higher gate fidelities are attainable through improving lattice spatial uniformity and reducing technical noises. 
Mid-circuit measurements can be realized by involving more internal states of orbit qubits, accompanied with spin-dependent transporting atoms to readout zones. 
Topological fault tolerance is also possible by further generating 3D cluster states \cite{raussendorf2007topological}. 
Moreover, integrating accordion lattices alongside advanced addressing techniques enables more precise local operations \cite{su2023dipolar}. 
Assembling approaches of continuously loading Bose-Einstein condensations will significantly increase the repetition rate of experimental cycles \cite{chen2022continuous}.

\begin{acknowledgments}

This work was supported by NNSFC Grant No. 12125409, Innovation Program for Quantum Science and Technology 2021ZD0302000.
W.-Y.Z.~acknowledges support from NSFC under Grant No. 12404323, the Fundamental Research Funds for the Central Universities under Grant No. WK2030000104, and the Postdoctoral Fellowship Program of CPSF under Grant No. GZC20241659 and No. 2025T180935.

\end{acknowledgments}



    
\bibliography{main}
\bibliographystyle{naturemag}

\onecolumngrid
\vspace*{0.5cm}
\newpage
\begin{center}
    \textbf{METHODS AND SUPPLEMENTAL MATERIALS}
\end{center}
\vspace*{0.5cm}

\twocolumngrid
\incltocpage
\tableofcontents
\appendix
\setcounter{secnumdepth}{2}

\twocolumngrid
\setcounter{equation}{0}
\setcounter{figure}{0}
\makeatletter
\makeatother
\renewcommand{\theequation}{S\arabic{equation}}
\renewcommand{\figurename}{Fig.}
\renewcommand{\thefigure}{S\arabic{figure}}
\renewcommand{\thetable}{S\arabic{table}}

\section{Experimental platform and superlattice structure}

As detailed in previous works, our optical lattice processor begins by preparing a two-dimensional gas of ultracold $^{87}$Rb atoms confined to a single antinode of a vertical lattice created by a $1064~\mathrm{nm}$ laser (labeled as the $z$-direction). 
After further evaporation, these atoms are loaded into a square lattice in the horizontal plane to form a nearly defect-free atom array spanning over 200 sites \cite{zhang2023scalable}, which serves as the starting point for our subsequent experiments. 
Along each in-plane axis (labeled as the $x$- and $y$-directions), we superimpose a blue-detuned ``short'' lattice ($\lambda_{\mathrm{s}} = 532~\mathrm{nm}$) and a red-detuned ``long'' lattice ($\lambda_{\mathrm{l}} = 1064~\mathrm{nm}$). 
A 50-degree, cross-angled incident design creates a lattice spacing of $a_\mathrm{s} = 630~\mathrm{nm}$ and $a_\mathrm{l} = 1260~\mathrm{nm}$ for the short and long lattice potentials, respectively. 

The superposition of both lattice potentials forms a tunable optical superlattice heterostructure along each in-plane axis
\begin{align}
V_{x(y)} = V_{\mathrm{s},x(y)} \cos^2(k_{\mathrm{s}} x(y)) - V_{\mathrm{l},x(y)} \cos^2(k_{\mathrm{l}} x(y) + \theta_{x(y)}),
\end{align} 
where $V_{\mathrm{s},x(y)}$ and $V_{\mathrm{l},x(y)}$ are the trap depths of the short lattice and the long lattice along the $x(y)$-direction, respectively. 
$k_{\mathrm{s}} = \pi / a_{\mathrm{s}}$ ($k_{\mathrm{l}} = \pi / a_{\mathrm{l}}$) is the wavenumber of the short lattice (long lattice). 
The relative phase $\theta_{x(y)}$ between the short lattice and the long lattice along the $x(y)$-direction is tuned with galvo-mount glass plates. 
Choosing $\theta_{x(y)} = \pi/4$ aligns the potential minima of two lattices, creating a staggered superlattice potential whose staggered offset is controlled solely by the long lattice depth $V_{\mathrm{l},x(y)}$.

Furthermore, applying a linear magnetic gradient adds a uniform tilt $\Delta_{x(y)}$ between neighboring sites across the lattice. 
When $\Delta_{x(y)}$ matches the staggered offset $\delta_{x(y)}$ ($\delta_{x(y)} \gg J_{x(y)}$), adjacent sites of opposite staggered offset form effective double-well potentials that isolate single-atom tunneling between neighboring double wells.
Orbit qubits are encoded in the superposition of left-right occupation within these double wells, as described in the main text. 
In principle, balanced superlattice potentials with $\theta_{x(y)} = 0$, used previously for creating multipartite spin-qubit entanglement \cite{zhang2023scalable}, could also hold the possibility for encoding orbit qubits. 
However, the staggered configuration used in this work offers several advantages.

Firstly, to realized isolated double wells in the staggered configuration (with $\theta_{x(y)} = \pi/4$), we only choose the long lattice depth so that the staggered offset satisfies $\delta_{x(y)} = \Delta_{x(y)}$, thereby producing detuning on the order of $h \times 1~\mathrm{kHz}$ (see main text and ``\textit{Calibration of parameters}'' \rm{section}), where $h$ is the Planck's constant. 
In contrast, in a balanced superlattice (with $\theta_{x(y)} = 0$), the long lattice potential must be several tens of times larger to achieve comparable suppression of inter-well tunneling. 
At this point, it is clear that the staggered configuration should be less sensitive to the inhomogeneities originating from the overall envelope of long lattice beams, and also more resilient to the fluctuations in relative phase $\theta_{x(y)}$ of the same strength.

Secondly, the additional linear magnetic gradient introduces a spin-dependent uniform tilt for different internal hyperfine states due to the differences in Land\'e $g$-factors and magnetic moments. 
For example, in our platform and in this work, we use $^{87}$Rb atoms and select two internal hyperfine states, $\ket{\downarrow} = \ket{F=1,m_F=-1}$ and $\ket{\uparrow} = \ket{F=2,m_F=-2}$.
The corresponding Land\'e $g$-factors are $-1/2$ and $1/2$, respectively. 
Consequently, assuming applying a linear magnetic gradient along the $x$-direction, the site-to-site energy offset increases by $|\Delta_{x,\downarrow}|=|\Delta_x|$ for $\ket{\downarrow}$ atoms but decreases by $|\Delta_{x,\uparrow}| = 2\Delta_x$ for $\ket{\uparrow}$ atoms.
This intrinsic, spin-dependent detuning enables the selective addressability and parallel control, which are essential for preparing and detecting cluster states with such a staggered orbit qubit scheme (see details in the main text), and are difficult to achieve in a balanced superlattice.

\section{Calibration of parameters}

\subsection{Parameters of lattice potential and magnetic gradient}

To find the resonant staggered potential $\delta_x = |\Delta_{x,\downarrow(\uparrow)}|$, we initialize the internal states of atoms to be $\ket{\downarrow} (\ket{\uparrow})$ and switch on the tunneling along $x$-direction. 
We fix the tunneling time to be $13.5~\mathrm{ms}$ (approximately $1.5$ times of the oscillation period) and measure the final imbalance dependent on the scanning $\delta_x$. 
For resonant $\delta_x$, most of the atoms tunnel to the opposite sites. 

After finding the resonant point, we can realize the resonant single-atom tunneling process ($R_Y$ gate) in effective double wells and extract the tunneling amplitude by fitting the oscillation in Fig.~\ref{fig2:Gate}a. 
The tunneling amplitudes for $\ket{\downarrow}$ and $\ket{\uparrow}$ are $J_{x,\downarrow}=58.00(5)~\mathrm{Hz}$ and $J_{x,\uparrow}=56.91(6)~\mathrm{Hz}$, respectively.

The magnetic gradient can be calibrated by the Ramsey process ($R_Z$ gate). 
As shown in Fig.~\ref{fig2:Gate}, after rotating the orbit qubits to the $x$-axis, we quench the $\delta_x$ to zero and hold for different periods of time $T$ before the final $\hat{R}_Y(\pi/2)$ gate. 
Then, the magnetic gradient can be extracted from the oscillation frequency. 
As shown in the main text, the magnetic gradients for $\ket{\downarrow}$ and $\ket{\uparrow}$ are $|\Delta_{x,\downarrow}|=954(3)~\mathrm{Hz}$ and $|\Delta_{x,\uparrow}|=1893(2)~\mathrm{Hz}$, respectively. 
The parameters along the $y$-direction can be calibrated by the same process. 
In this work, the magnetic gradients along $y$ direction are $|\Delta_{y,\downarrow}|=993(4)~\mathrm{Hz}$ and $|\Delta_{y,\uparrow}|=1940(5)~\mathrm{Hz}$.

For calibrating on-site interaction, we initialize the spin and density configuration as Fig.~\ref{fig1:Encoding}c left and switch on the tunneling along $y$-direction ($J_{y,\downarrow}=58.6(4)~\mathrm{Hz}$). 
We scan the $\delta_y$ under a fixed tunneling time of $4.3~\mathrm{ms}$ (approximately half the oscillation period). 
When $|\delta_y-\Delta_{y,\downarrow}| \approx U$, most of the $\ket{\downarrow}$ atoms tunnel to opposite sites and can be detected by our density-resolved measurement technique \cite{wang2023interrelated,zhang2024observation}. 
In this work, $U=447(2)~\mathrm{Hz}$.

\subsection{Coherence of orbit qubits}
\begin{figure*}[t!]
  \centering
  \includegraphics[width=1\textwidth]{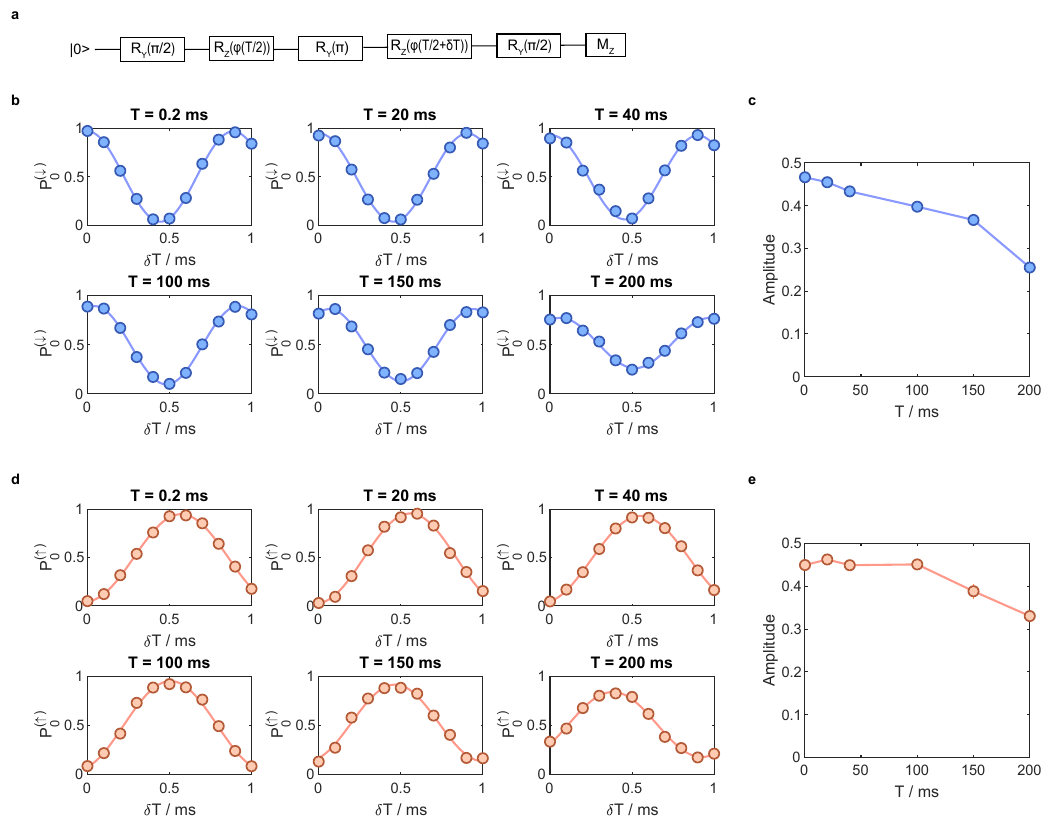}
  \caption{\textbf{Coherence of orbit qubits.}
  \textbf{a}, Quantum circuit to detect the coherence of orbit qubits.
  \textbf{b}, The coherent oscillation for $\ket{\downarrow}$ qubit dependent on $\delta T$ for different hold time $T$.
  \textbf{c}, The fitted oscillation amplitudes in \textbf{b} dependent on $T$.
  \textbf{d}, The coherent oscillation for $\ket{\uparrow}$ qubit dependent on $\delta T$ for different hold time $T$.
  \textbf{e}, The fitted oscillation amplitudes in \textbf{d} dependent on $T$.}
  \label{figS1:Coherence}
\end{figure*}

Note that idle orbit qubits can experience different $z$-rotations due to the inhomogeneity of chemical potential. 
To verify the maintenance of coherence for orbit qubits, we employ a spin-echo sequence, as shown in Fig.~\ref{figS1:Coherence}a. 
We introduce additional tilts in effective double wells during hold time, and the slight difference of time $\delta T$ between pulses leads to effective $R_Z$ rotation on orbit qubits. 
Fig.~\ref{figS1:Coherence}a,c show the oscillation of the final location dependent on $\delta T$. 
After fitting the oscillation for different $T$, we show the decay of amplitudes in Fig.~\ref{figS1:Coherence}b,d. 
The lifetimes are both larger than $200~\mathrm{ms}$, and the oscillation amplitude decays little until $40~\mathrm{ms}$.

\section{Generation and detection of cluster states}

\subsection{Detecting multipartite entanglement}
In our previous work \cite{zhang2023scalable}, we have provided mathematical definitions of the two terminologies \textit{full bipartite non-separability} and \textit{genuine multipartite entanglement}. 
In short, a quantum state possesses \textit{full bipartite non-separability} if it cannot be expressed as
\begin{align}
\rho = \sum_{i}{p_i \ket{\psi^{(i)}_{M}} \bra{\psi^{(i)}_{M}} \otimes \ket{\psi^{(i)}_{\bar{M}}}\bra{\psi^{(i)}_{\bar{M}}}}
\end{align}
for arbitrary bipartition $M|\bar{M}$ of the qubits in the quantum state. 
A quantum state is a \textit{genuine multipartite entanglement} if it cannot be expressed as
\begin{align}
\rho = \sum_i{p_i\ket{\psi^{(i)}_{M_i}}\bra{\psi^{(i)}_{M_i}}\otimes \ket{\psi^{(i)}_{\bar{M}_i}}\bra{\psi^{(i)}_{\bar{M}_i}}},
\label{bs}
\end{align}
where the bipartition $M_i|\bar{M}_i$ can be different for different $i$. 
It is clear that \textit{genuine multipartite entanglement} is a stronger claim than \textit{full bipartite non-separability}.

The bipartite entanglement witness (\ref{EW1D}) and (\ref{EW2D}) detect the entanglement between adjacent qubits, which can be directly proved by the Cauchy-Schwarz inequality and the anticommutativity theorem. 
For a quantum state consisting of $N$ qubits, we can generate a graph where each vertex corresponds to a qubit, and qubits $\{i,j\}$ are connected if they are entangled. 
If it is a connected graph, the quantum state is $N$-qubit \textit{full bipartite non-separability}. 
As shown in Fig.~\ref{fig3:Cluster}c and Fig.~\ref{fig4:Cluster}c, we have verified \textit{full bipartite non-separability} for 1D and 2D cluster states.

To detect the \textit{genuine multipartite entanglement} for the 1D case, we adopt the GME witness (\ref{EWGME0}). This witness can be proved as follows. 
Consider a bi-separable pure state $\ket{\phi} = \ket{\phi_M}\ket{\phi_{\bar{M}}}$, where $i\in M$ and $i+1 \in \bar{M}$. Then
\begin{align}
\langle \hat{W}^{(\mathrm{GME})}_{L}\rangle \geq \langle \hat{W_i}\rangle \geq 0.
\end{align}
Since $\langle \hat{W}^{(\mathrm{GME})}_{L}\rangle \geq 0$ holds for arbitrary bipartition $M|\bar{M}$, the states that can be expressed in Eqn.(\ref{bs}) also fulfills $\langle \hat{W}^{(\mathrm{GME})}_{L}\rangle \geq 0$. 
Equivalently, the 1D chain is GME if $\langle \hat{W}^{(\mathrm{GME})}_{L}\rangle < 0$.

\subsection{Realization of required local measurement settings}
To verify the cluster states, we measure the stabilizers under the local measurement settings shown in Fig.~\ref{fig3:Cluster}a and Fig.~\ref{fig4:Cluster}a. 
According to the arrangement of spin-dependent orbit qubits, these settings can be realized by applying selective $\hat{R}_Y(\pi/2)$ operations for one type of internal state before location readout.

To compensate for the additional $R_Z$ rotation in the circuits during idling and mitigate the spatial inhomogeneity effect, we again employ a spin-echo sequence. 
After the preparation of cluster states, we apply a $\hat{R}_Y(\pi)$ gate on $\ket{\uparrow}$ or $\ket{\downarrow}$ qubits, and hold for a specific period of time before the final $\hat{R}_Y(\pi/2)$ operation. During the hold time, we vary the staggered potential $\delta_x,\delta_y$ and keep $J_x, J_y=0$.

During the initialization of orbit qubits, we apply $\hat{R}_Y(\pi/2)$ gates to the $\ket{\downarrow}$ and $\ket{\uparrow}$ qubits in sequence. 
The order leads to different conditions for compensation of the additional $R_Y$ rotation. 
For the $X$ measurement on $\ket{\uparrow}$ qubits, the compensation can be directly realized by reversing the staggered potential $\delta_x,\delta_y$ accordingly. 
As for $\ket{\downarrow}$ qubits, the short lattice along $x$-direction is shallow during $R_Y^{(\uparrow)}$, leading to a change of Wannier function and an effective $z$-rotation different to the case of deep lattice. 
Thereby, in addition to reversing the staggered potential, we slightly modify the hold time. 

The spin-echo sequence enables the simultaneous measurement of stabilizers of cluster states in a large region.

\subsection{Mitigation of measurement errors}
We calibrate the $X$ measurement error from the spin-echo sequence, which can be regarded as two consecutive times of $X$ measurement steps. 
Assume that the probability of returning to the initial state after a spin-echo is $p$. 
We deduce the accuracy of $X$ measurement is $\sqrt{p}$. 
Let $S~(\widetilde{S})$ denotes an observable containing one $X$ operator measured with the ideal (noisy) $X$ measurement process and define $S=2P_S-1~(\widetilde{S}=2\widetilde{P}_S-1)$.
We have
\begin{align}
\widetilde{P}_S &= P_S\sqrt{p}+(1-P_S)(1-\sqrt{p}) \\ \notag
&=(2\sqrt{p}-1)P_S+(1-\sqrt{p}).
\end{align}
So,
\begin{align}
\widetilde{S} &= 2\widetilde{P}_S-1=2((2\sqrt{p}-1)P_S+(1-\sqrt{p}))-1 \\ \notag
&=(2\sqrt{p}-1)(2P_S-1)=(2\sqrt{p}-1)S.
\end{align}
By measuring the spatial distribution of $p$, we correct the stabilizer measurements in Fig.~\ref{fig3:Cluster} and Fig.~\ref{fig4:Cluster} according to the relation above. 
For Fig.~\ref{fig3:Cluster}b, the average $p$ in the black rectangles is $p=96.76(8)\%$. 
For Fig.~\ref{fig4:Cluster}b, the average $p$ in the black rectangles is $p=93.75(12)\%$. 
The lower value in the 2D case is due to the longer time for the quantum circuit.

\subsection{Extracting fidelity of CZ gates}

To extract the fidelity of the CZ gates, we need to correct both the initialization and measurement errors. 
Since complete entanglement circuits contain the spin-echo sequence and CZ gates, the average fidelity of CZ gates can be extracted from 1D stabilizer measurements by
\begin{align}
F_i = \sqrt{\frac{\langle\hat{g}_i\rangle}{\langle\hat{X}_i\rangle_{\mathrm{spin-echo}}}}.
\end{align}
Here subscript $i$ labels the index of qubit and $\langle\hat{X}_i\rangle_{\mathrm{spin-echo}}$ corresponds to results of $X$ measurement after single-qubit initialization in the spin-echo sequence. 
In the purple rectangle region in Fig.~\ref{fig3:Cluster}c, the average fidelity is $94.8(13)\%$.

\subsection{Stabilizer map of 1D and 2D cluster states}
\begin{figure*}[t!]
  \centering
  \includegraphics[width=1\textwidth]{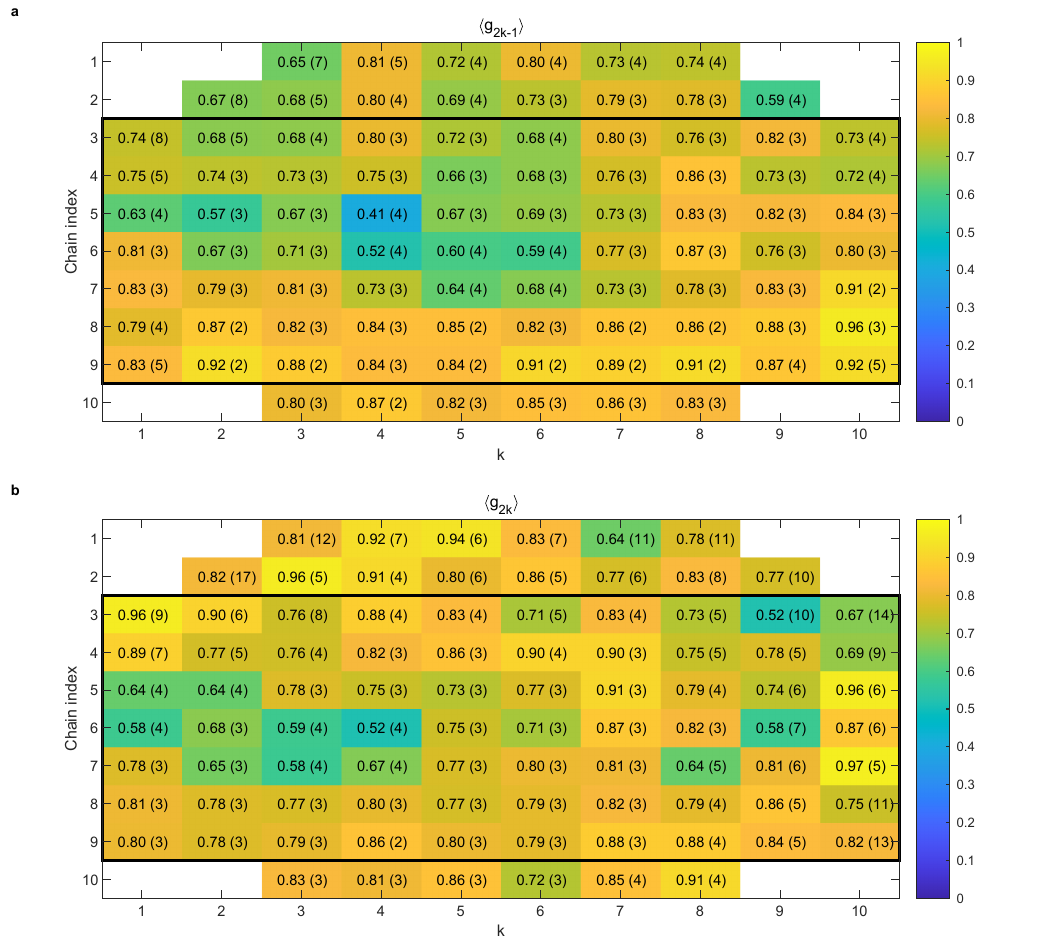}
  \caption{\textbf{Map of 1D stabilizers.}
  Detailed data of Fig.~\ref{fig3:Cluster}b in the main text. Within the region marked by black rectangles, the average values of the stabilizer $\langle \hat{g}_i \rangle$ for odd and even $i$ are $0.769(1)$ (top) and $0.775(1)$ (bottom), respectively. The values here have been corrected to mitigate the measurement errors.}
  \label{figS2:1DMap}
\end{figure*}

\begin{figure*}[t!]
  \centering
  \includegraphics[width=1\textwidth]{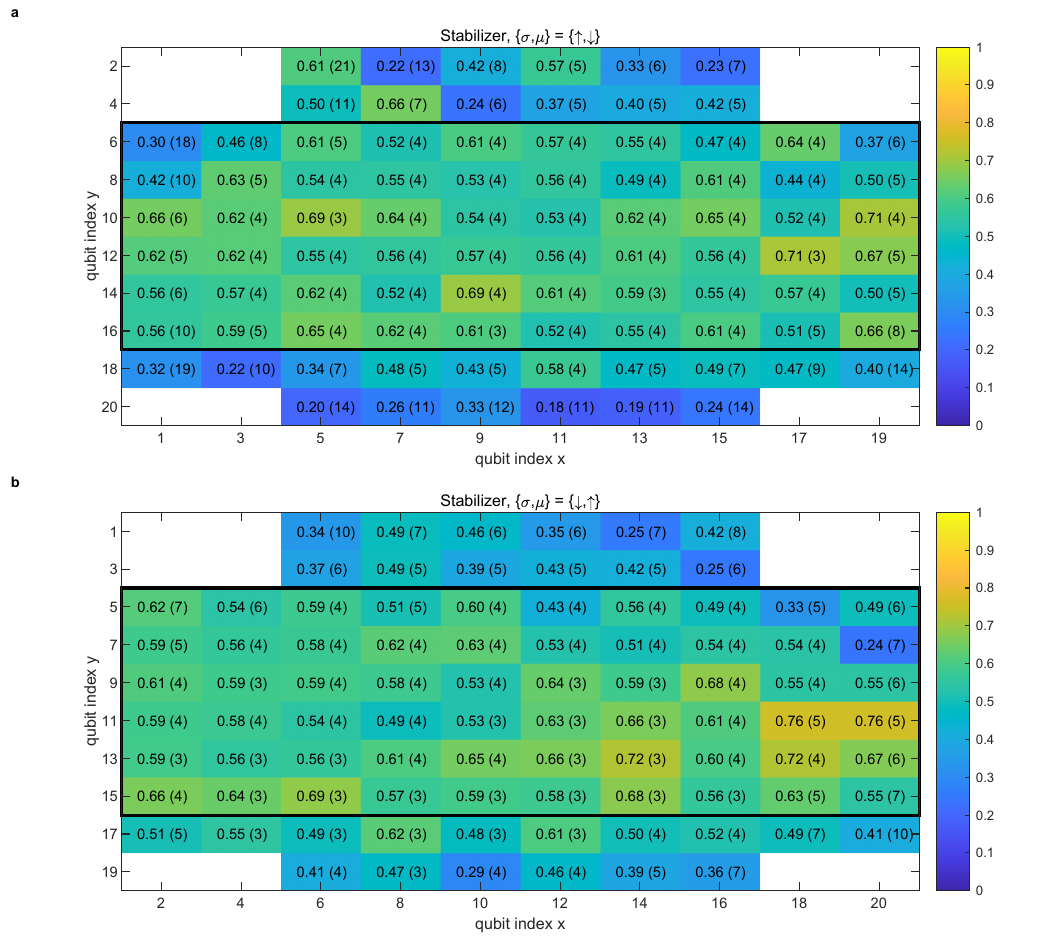}
  \caption{\textbf{Map of 2D stabilizers.}
  Detailed data of Fig.~\ref{fig4:Cluster}b in the main text. Within the region marked by black rectangles, the average stabilizer values are $0.572(7)$ (top) and $0.584(6)$ (bottom), respectively. The values here have been corrected to mitigate the measurement errors.}
  \label{figS3:2DMap}
\end{figure*}

Fig.~\ref{figS2:1DMap} and Fig.~\ref{figS3:2DMap} show the map of measured stabilizers, which have been corrected for measurement errors. 
In experiments, there exist empty double wells in each shot, due to the imperfect initial filling, leakage in quantum circuits, and atom loss during measurements. 
To mitigate the error above, for each $m$-qubit stabilizer, we locally post-select the density configurations that contain one atom in each supported double well.

\section{Demonstration of measurement-based logical gates on 1D cluster states}

\subsection{Basic theory}
The principle of realizing logical single-qubit gates under the measurement-based quantum computation framework can be illustrated in Fig.~\ref{fig5:MBQC}a.
After entangling a qubit in $\ket{\psi}$ (the `input' state) with a qubit in $\ket{+}$, measuring the first qubit in the basis of
\begin{align*}
\mathcal{B_{\theta}} = \left\{\frac{\ket{0}+e^{i\theta}\ket{1}}{\sqrt{2}},\frac{\ket{0}-e^{i\theta}\ket{1}}{\sqrt{2}}\right\}
\end{align*}
leads to an `output' state $\hat{U}_s(\theta)\ket{\psi}$ dependent on the measurement outcome $s$ ($s = \pm 1$). Here
\begin{align*}
\hat{U}_s(\theta) &= \left[\left(
\begin{matrix}
\frac{1}{\sqrt{2}},\frac{(-1)^se^{-i\theta}}{\sqrt{2}}
\end{matrix}
\right)\otimes \hat{I}\right]
CZ
\left[\hat{I}\otimes\left(
\begin{matrix}
\frac{1}{\sqrt{2}}\\
\frac{1}{\sqrt{2}}
\end{matrix}
\right)\right]\\
&\propto \hat{H}\hat{Z}^s\hat{R}_Z(\theta),
\end{align*}
with
\begin{align*}
\hat{R}_Z(\theta) = \left(
\begin{matrix}
1 & 0\\
0 & e^{-i\theta}
\end{matrix}
\right).
\end{align*}

Based on the above-mentioned principle, we can demonstrate logical gates on the prepared 1D cluster states. 
Fig.~\ref{fig5:MBQC} shows local measurement settings for different logical-qubit gates and the corresponding teleportation of quantum states. 
All of the `input' states are initialized to $\ket{+}$. 
In the following, we further provide the theoretical details of the realized logical gates.

In Fig.~\ref{fig5:MBQC}b, $\theta_1=\theta_2=0$, so
\begin{align*}
\ket{\psi_{\mathrm{output}}} = (\hat{H}\hat{Z}^{s_2})(\hat{H}\hat{Z}^{s_1})\ket{\psi_{\mathrm{input}}} = \hat{X}^{s_2}\hat{Z}^{s_1}\ket{\psi_{\mathrm{input}}}.
\end{align*}
The only effect of the random byproduct operator $\hat{X}^{s_2}\hat{Z}^{s_1}$ is that we need to reinterpret the final readout measurement.
So, the evolution operator $\hat{U}(s_1,s_2)=\hat{X}^{s_2}\hat{Z}^{s_1}$ is equivalent to an identity gate.
For $\ket{\psi_{\mathrm{input}}}=\ket{+}$, the output state is $\ket{\psi_{\mathrm{output}}}\propto (\ket{0}+(-1)^{s_1}\ket{1})/\sqrt{2}$,
which fulfills $\langle \hat{X}\rangle = (-1)^{s_1}$.

In Fig.~\ref{fig5:MBQC}c, $\theta_1=\pi/2, \theta_2=\eta$, so
\begin{align*}
\ket{\psi_{\mathrm{output}}} &= [\hat{H}\hat{Z}^{s_2}\hat{R}_Z(\eta)][\hat{H}\hat{Z}^{s_1}\hat{R}_Z(\pi /2)]\ket{\psi_{\mathrm{input}}}\notag \\
&= (\hat{X}^{s_2}\hat{Z}^{s_1})\hat{R}_X((-1)^{s_1} \eta)\hat{R}_Z(\pi /2)\ket{\psi_{\mathrm{input}}}.
\end{align*}
The evolution operator $(X^{s_2}Z^{s_1})\hat{R}_X((-1)^{s_1} \eta)\hat{R}_Z(\pi /2)$ is equivalent to a rotation gate.
For $\ket{\psi_{\mathrm{input}}}=\ket{+}$,
the output state fulfills $\langle \hat{Y}\rangle = (-1)^{s_1+s_2+1}\cos{\eta}$.

In Fig.~\ref{fig5:MBQC}d, $\theta_1=\theta_2=\theta_3=0$, so
\begin{align*}
\ket{\psi_{\mathrm{output}}} &= (\hat{H}\hat{Z}^{s_3})(\hat{H}\hat{Z}^{s_2})(\hat{H}\hat{Z}^{s_1})\ket{\psi_{\mathrm{input}}}\notag \\
&\propto (\hat{X}^{s_1+s_3}\hat{Z}^{s_2})\hat{H}\ket{\psi_{\mathrm{input}}}.
\end{align*}
The evolution operator $\hat{U}(s_1,s_2,s_3)=(\hat{X}^{s_1+s_3}\hat{Z}^{s_2})\hat{H}$ is equivalent to a Hadamard gate.
For $\ket{\psi_{\mathrm{input}}}=\ket{+}$, the output state is
$\ket{\psi_{\mathrm{output}}} = \ket{s_1+s_3}$,
which fulfills $\langle \hat{Z}\rangle = (-1)^{s_1+s_3}$.

Fig.~\ref{fig5:MBQC}e shows the measurement setting to demonstrate a logical entangling gate. 
The first and fourth qubits correspond to the input qubits. 
After measurement on qubits $\{1,4\}$, the quantum information propagates to qubits $\{2,3\}$, which are logically entangled with a CZ gate. 
The output state fulfills
\begin{align*}
&~~~~\ket{\Psi_{\mathrm{output}}} \notag \\
&= \hat{U}_{CZ}\left\{[\hat{H}\hat{Z}^{s_1}\hat{R}_Z(\pi /2)]\otimes[\hat{H}\hat{Z}^{s_4}\hat{R}_Z(\pi /2)]\right\}\ket{\Psi_{\mathrm{input}}}.
\end{align*}
For $\ket{\Psi_{\mathrm{input}}} = \ket{+}\ket{+}$, the output state is
\begin{align*}
\ket{\Psi_{\mathrm{output}}} = \hat{U}_{CZ}\left[\frac{\ket{0}+i(-1)^{s_1}\ket{1}}{\sqrt{2}}\otimes \frac{\ket{0}+i(-1)^{s_4}\ket{1}}{\sqrt{2}}\right],
\end{align*}
with stabilizers $(-1)^{s_1}\hat{Y}_2\hat{Z}_3$ and $(-1)^{s_4}\hat{Z}_2\hat{Y}_3$.

\subsection{Experimental realization of logical quantum gates}
\begin{figure*}[t!]
  \centering
  \includegraphics[width=1\textwidth]{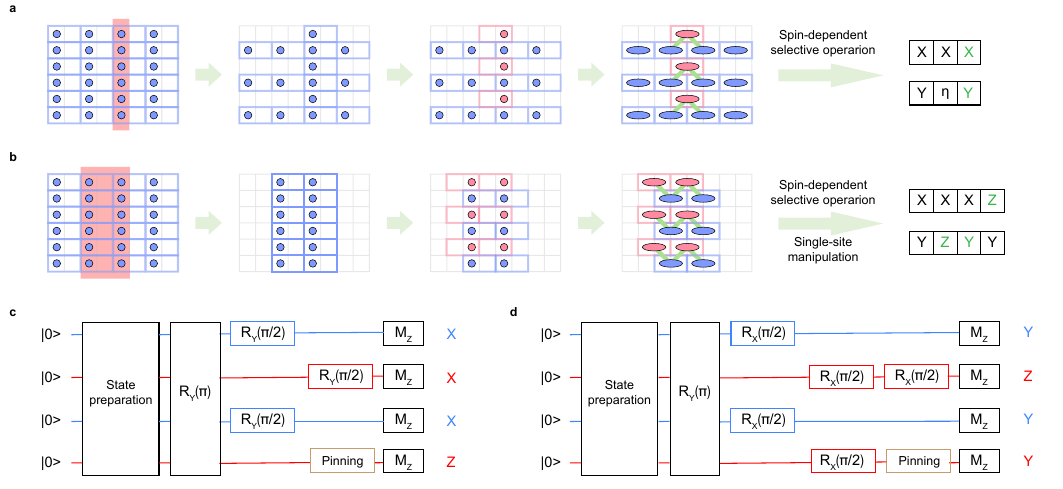}
  \caption{\textbf{Measurement-based quantum computing based on orbit qubits.}
  \textbf{a}, \textbf{b}, Schematic of the process to generate target entangled chains with specific length. We demonstrate logical identity gate and rotation gate with length-$3$ cluster states (in \textbf{a}, realizing local measurement settings with only spin-selective operation. To demonstrate logical Hadamard gate and entangling gate, we use length-$4$ cluster states (in \textbf{b}) and realize corresponding settings with additional local manipulation.
  \textbf{c}, Quantum circuit for demonstrating logical Hadamard gate.
  \textbf{d}, Quantum circuit for demonstrating a logical entangling gate.}
  \label{figS4:MBQC}
\end{figure*}

We prepare orbit-qubit chains with a specific length by leveraging single-site addressing and spin-dependent superlattice techniques, which have been demonstrated in our previous works. 
Fig.~\ref{figS4:MBQC}a,b show the process to generate target chains. 
We start from alternating columns of atoms in $\ket{F=1,m_F=-1}$, prepared by the staggered-immersion cooling method. 
Using the addressing laser (with a magic wavelength of $787.55~\mathrm{nm}$ and $\sigma^{+}$ polarization) to project specific patterns with a digital micromirror device (DMD), we can shift the resonant frequency between $\ket{\uparrow}$ and $\ket{\downarrow}$ for selected atoms (marked with red shadows in Fig.~\ref{figS4:MBQC}a,b left). 
In Fig.~\ref{figS4:MBQC}a, we switch on spin-dependent superlattice along $y$-direction in addition, and flip the atoms in odd rows (except for the addressed atoms). 
In Fig.~\ref{figS4:MBQC}b, we flip all the atoms outside the marked region. 
After sweeping the atoms in $\ket{\uparrow}$, we can obtain specific density configurations. 
Then, we flip the spin of atoms in odd rows and follow the same process to generate 1D cluster states, preparing 1D orbit-qubit chains with length 3 (Fig.~\ref{figS4:MBQC}a) and length 4 (Fig.~\ref{figS4:MBQC}b).

The measurement settings in Fig.~\ref{figS4:MBQC}a (corresponding to the identity gate and the rotation gate) can be directly realized by spin-selective operation.
For the measurement settings in Fig.~\ref{figS4:MBQC}b (related to the $\hat{H}$ gate and the entangling gate), we also use the single-site manipulation.
Fig.~\ref{figS4:MBQC}c,d show the quantum circuits to realize these measurement settings. 
In the circuits, the colors of the lines represent the internal states of orbit qubits (red for $\ket{\uparrow}$ and blue for $\ket{\downarrow}$). 
The $\hat{R}_Y(\pi)$ gates after state preparation are used for the spin echo sequence. 
Single-qubit gates labeled by red and blue rectangles are performed with spin-dependent selective manipulation. 
The `Pining' process is realized by projecting a local repulsive potential to freeze the tunneling dynamics of the corresponding orbit qubits.

In Fig.~\ref{figS4:MBQC}c, the fourth qubit is pinned during the final $\pi$-pulse for $\ket{\uparrow}$ qubits. 
This leads to an effective $Z$ measurement for the final qubit and $X$ measurements for other qubits.
In fact, based on multi-step global operation and local pinning, an arbitrary local measurement can be realized. 
For example, if a local measurement for two qubits requires operation $\hat{U}_{1}\otimes\hat{U}_2$ before the final readout, we can globally perform $\hat{U}_1$ and $\hat{U}_{2}\hat{U}^{-1}_{1}$ in sequence, pinning the first qubit during the second step. 
We employ this alternating protocol to demonstrate the logical entangling gate. 
Our result shows that a robust local pinning process is enough for arbitrary local measurement settings.

In the future, more internal states can be used. 
By appropriately initializing the spin configuration, a simple local measurement can also be realized by a selective operation directly. 
Our spin-dependent orbit-qubit processor has the potential to strongly reduce the overhead of local operations for MBQC.

\end{document}